\documentclass[aps,pre,twocolumn,groupedaddress,showpacs,floatfix]{revtex4}
\usepackage{amsmath}
\usepackage{amssymb}
\usepackage{graphicx}

\begin{document}

\title{Complex Networks, Simple Vision}

\author{Luciano da Fontoura Costa}
\affiliation{Institute of Physics of S\~ao Carlos. 
University of S\~ ao Paulo, S\~{a}o Carlos,
SP, PO Box 369, 13560-970, 
phone +55 162 73 9858,FAX +55 162 73
9879, Brazil, luciano@if.sc.usp.br}

\date{March 5, 2004}

\begin{abstract}   

This paper proposes and illustrates a general framework to integrate
the areas of vision research and complex networks.  Each image pixel
is associated to a network node and the Euclidean distance between the
visual properties (e.g. gray-level intensity, color or texture) at
each possible pair of pixels is assigned as the respective edge
weight.  In addition to investigating the therefore obtained weight
and adjacency matrices in terms of node degree densities, it is shown
that the combination of the concepts of network hub and
\emph{2-}expansion of the adjacency matrix provides an effective means
to separate the image elements, a challenging task in computer vision
known as \emph{segmentation}.

\end{abstract}


\maketitle

\section{Introduction}

While effortless to humans and animals, the visual interpretation of
scenes -- i.e. the process of vision \cite{Marr}-- represents one of
the greatest challenges to science.  Indeed, no automated vision
system \cite{CostaCesar:2001} has so far been obtained whose
performance can be compared even to those of the simplest animals.  At
the same time, relatively little is known about how biological vision
operates.  Great part of the difficulty in understanding and
implementing vision derives from the lack of general concepts and
approaches to represent, characterize and model the intricate data
and processes underlying visual processing.  One particularly relevant
aspect is the fact that 2D images are normally obtained by projecting
3D scenes onto some planar surface, implying substantial loss of
information and degeneration.  One immediate effect of such
projections is the fact that the spatial adjacency relations are
abruptly changed, in the sense that two spatially separated objects in
the 3D scene may become adjacent in the 2D projection. At the same
time, the superposition of objects (i.e. occlusion) may break the
image background (or another object) into two or more disconnected
portions.  For instance, the pen in Figure~\ref{fig:Lapis}(a)
separates the white background into two parts.  Another interesting
aspect is the fact that, because of image resolution constraints
implied by the size of the photoreceptors, stereo vision becomes
useless for objects further away than about 100 meters.  Even so, the
fact that humans are still fully able to interpret such scenes and
print pictures clearly indicates that there are effective solutions to
the problem of understanding 2D images.  One of the hardest problems
in vision is the process of partitioning the image into meaningful
connected components, which is known as \emph{segmentation}
\cite{CostaCesar:2001,jain_clust:1999}.  Despite continuing attention
from the scientific community, the problem of finding a segmentation
approach capable of producing suitable results for general images
remains a considerable challenge, as a consequence of the
degenerations implied by 2D imaging allied to other problems such as
reflexes and shades.  While fully general and automated image
segmentation will ultimately depend on the incorporation of high-level
information about image formation and the objects to be expected in
image (i.e. good \emph{models} of the visual world), it is argued here
that reasonable segmentations can be obtained by considering the full
range of spatial interactions between the image elements.

The interesting problems implied by vision have motivated approaches
based on areas ranging from partial differential equations
(e.g. \cite{Sethian:1999}) to wavelets (e.g. \cite{Starck:1998}).
Several physics-based approaches to vision \cite{Shaffer:1992} have
been reported in the literature, including the use of the Potts model
for image segmentation (e.g. \cite{Potts_PNAS:2003}), deformable
contours (e.g. \cite{Blake:1998}, random fields (e.g. \cite{Li:1995})
and fractals (e.g. \cite{fractalbased:1996}).  While the projection of
2D scenes can be modeled as random scalar fields, digital images are
normally represented in terms of matrices, where each element
corresponds to a pixel, which provides a poor representation of the
original visual information in the sense that it does not incorporate
any information about spatial interactions and correlations along
several spatial ranges between the pixels.  Several graph-theoretical
approaches to image analysis and computer vision, interesting
graph-based approach to visual representation and analysis
\cite{shoko:2002} involve the use of graphs where each pixel is
associated to a node and the difference between the visual properties
of adjacent pixels are used to define the respective edge weights.
One particularly interesting graph-based method called image foresting
transform -- IFT -- has been described (e.g. \cite{Falcao:2000}) where
minimal paths are identified between pixels considering some spatial
adjacency, yielding a forest
\footnote{A \emph{forest} is an acyclic graph.} of minimum paths
connecting the image pixels.  Several useful image properties can be
derived from such graph representations, including the segmented image
regions \cite{Falcao:2000} and multiscale skeletons
\cite{Falcao:2002}.

The current work proposes a complex network \cite{Albert_Barab:2002,
Dorog_Mendes:2002, Newman:2003} approach to vision research where each
pixel is represented as a node and the distance between the visual
properties (e.g. gray-level, color or texture) of \emph{every} pair of
pixels in the image is adopted as the respective edge weight.  The
distance between the pixels can also be considered into the edge
weight.  The resulting fully-connected graph is subsequently
thresholded at a specific value $T$ in order to obtain the respective
adjacency matrix A \cite{Costa_Hier:2003}.  It is argued here that, by
incorporating all ranges of spatial interactions between the image
elements, such representations present good potential for integrating
the relevant visual properties from the lowest to the highest levels
of abstraction, leading to more effective visual processing.
Moreover, it is also suggested that the use of the concepts and tools
underlying complex network research (e.g. node degree, clustering
coefficient, shortest path, \emph{L-}expansions and critical
phenomena) can provide relevant information to be used for image and
object characterization.  Such a potential is illustrated in this
paper with respect to image characterization in terms of node degree
densities and the use of \emph{L-}expansions of the graph in order to
obtain enhanced image segmentation.

\section{Image Representation as a Complex Network}

The original gray-level image is assumed to have $M \times M$ pixels,
each of which can take gray-level values ranging from $0$ to $G$,
which is proportionally normalized into the interval $[0,1]$.
Therefore, the network representing the image contains $N=M^2$ pixels
and $n=N(N-1)/2$ weighted edges, which are represented by the $N
\times N$ weight matrix $W$. Equations~\ref{eq:map1}-~\ref{eq:map3}
define a possible mapping between each node $i$ of the network and the
image pixels $(x,y)$, where $1 \leq x,y \leq M$ and the function
$mod(a,b)$ stands for the remainder of $a$ divided by $b$.

\begin{eqnarray}
  i = y + (x-1)M \label{eq:map1}  \\
  x = \lfloor (k-1)/M \rfloor + 1 \label{eq:map2}  \\
  y = mod((k-1),M) + 1 \label{eq:map3}  
\end{eqnarray}

Several types of interactions between two pixels can be defined
respectively to each possible visual attribute, such as light
intensity, color components, local shape, texture, motion, and
disparity (stereo), as well as the pixels position and spatial
adjacency \cite{Costa_Vor:2003} between pairs of pixels.  The scalar
values derived from such properties can be organized into a feature
vector $\vec{f}$ in $R^N$, in such a way that each network node $i$
becomes associated to a feature vector $\vec{f}_i$ describing the
local visual properties around the respective image pixel.  The weight
of the edge connecting two nodes $i$ and $j$ is therefore defined as
$w(j,i) = \| \vec{f}_i -\vec{f}_j \|$, where $\| a \|$ denotes the
Euclidean norm.  As such, the edge weight reflects the visual
dissimilarity between the respective pair of pixels.  In case the
visual features take into account a small neighborhood around each
pixel, such as is the case with texture or motion, the edge weights
will quantify the dissimilarity between local properties around the
pairs of pixels.  Let $\delta_T(a)$ be the operator acting over the
matrix $a$ in such a way that the value one is assigned whenever the
respective element of $a$ has absolute value larger or equal than the
specified threshold $T$; e.g., $\delta_2(\vec{x}=(4 ,-2, 0, -3,
1))=(1,1,0,1,0)$.  Thus, the adjacency matrix $A$ underlying the
weighted complex network $\Gamma$ for a specific threshold value $T$
can be obtained as $A=\delta_T(W)$.  The average node degree of
$\Gamma$ is henceforth represented as $z$.

The concept of \emph{L-}expansion of a network, introduced in
\cite{Costa_exp:2003} provides an interesting possibility for
identifying \emph{L-}paths along the represented network.  The
\emph{L-}expansion of a network $\Gamma$ \cite{Costa_exp:2003}
consists of the network containing the same number of nodes as
$\Gamma$ where an edge is incorporated between nodes $i$ and $j$
whenever a path of length $L$ is found to exist between those nodes in
the original network $\Gamma$.  Because of the high computational cost
implied for the calculation of such expansions for large values of
$L$, the current work is limited to $L=2$, in which case the adjacency
matrix $A2$ of the \emph{2-}expanded network can be immediately
obtained from the squared adjacency matrix $B=A^2$ while ignoring the
elements in the resulting main diagonal, i.e.

\begin{equation}
  A2(i,j) = \left\{ \begin{array}{ll}
                      B(i,j)  & \mbox{if $i \neq j$} \\
                      0         & \mbox{otherwise}
                   \end{array}  \right.  
\end{equation}

\begin{figure}
 \begin{center} 
   \includegraphics[scale=.55,angle=-90]{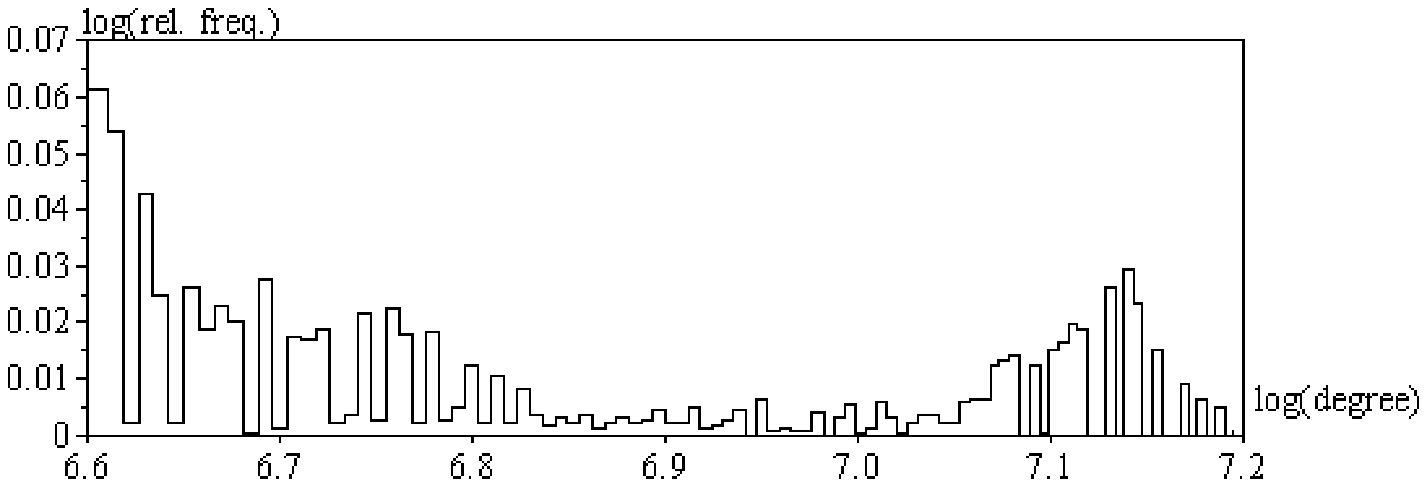} \\
   (a) \\
   \includegraphics[scale=.55,angle=-90]{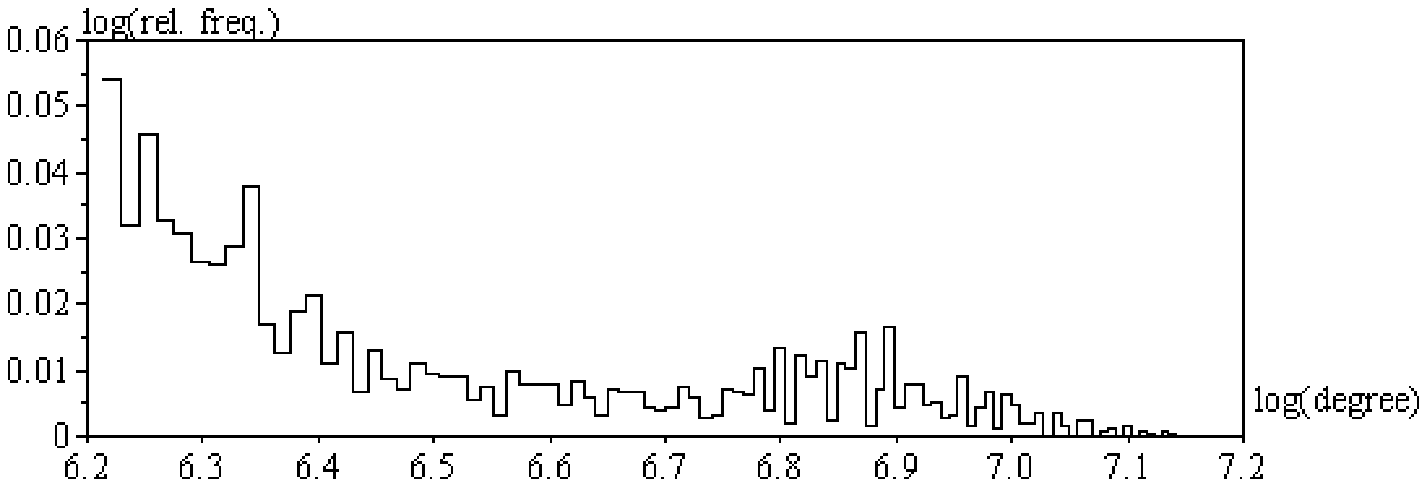} \\
   (b) \\   
   \includegraphics[scale=.55,angle=-90]{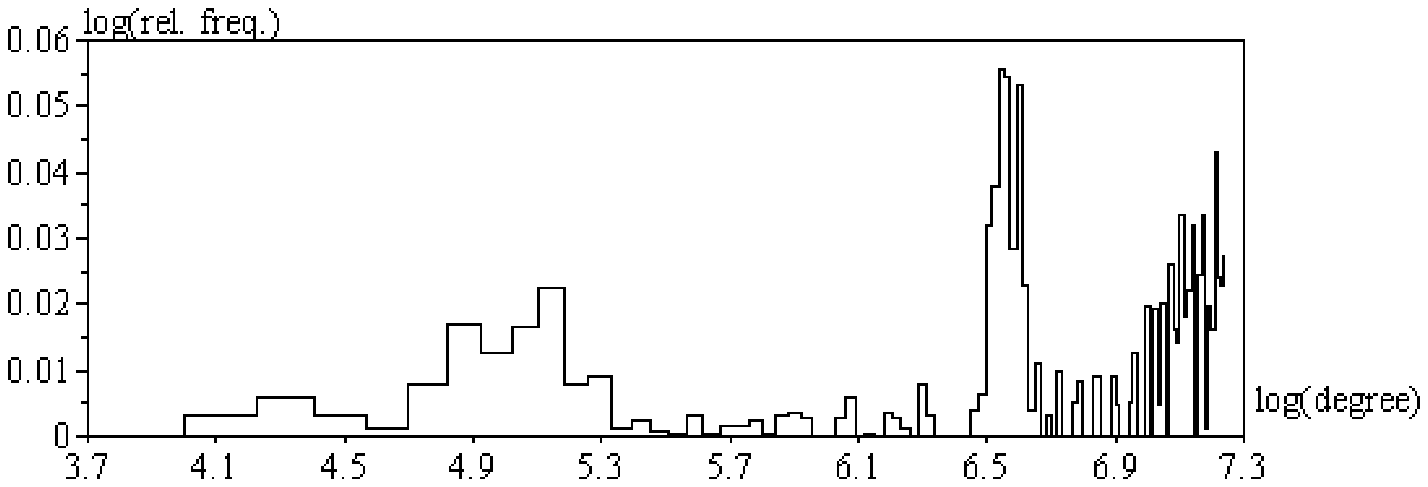}  \\
   (c) \\
   \includegraphics[scale=.55,angle=-90]{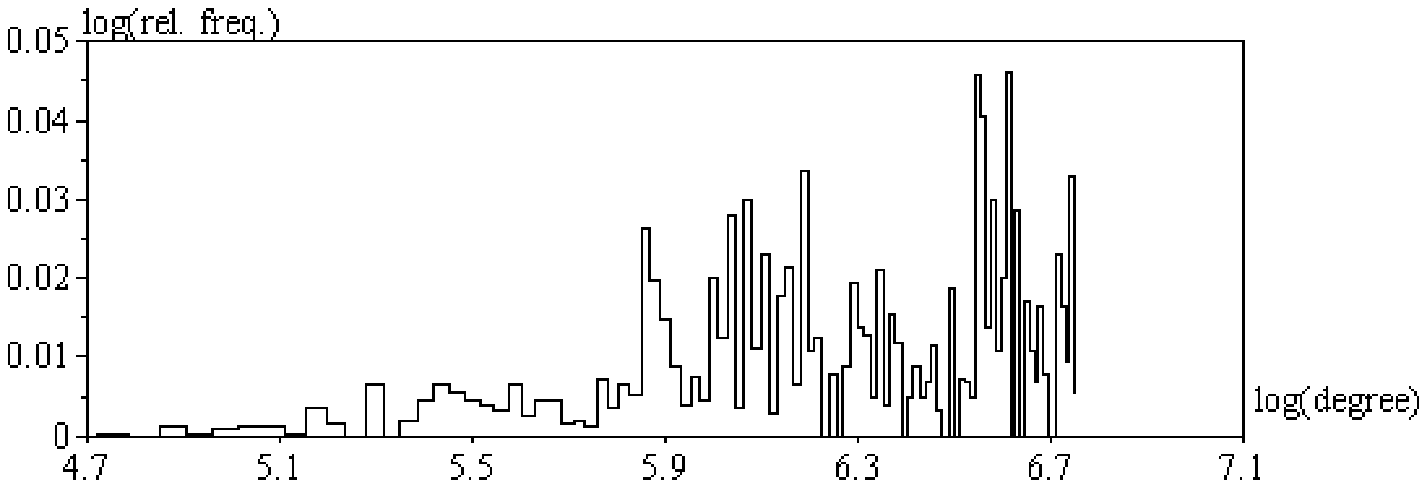} \\
   (d) \\   
   \includegraphics[scale=.55,angle=-90]{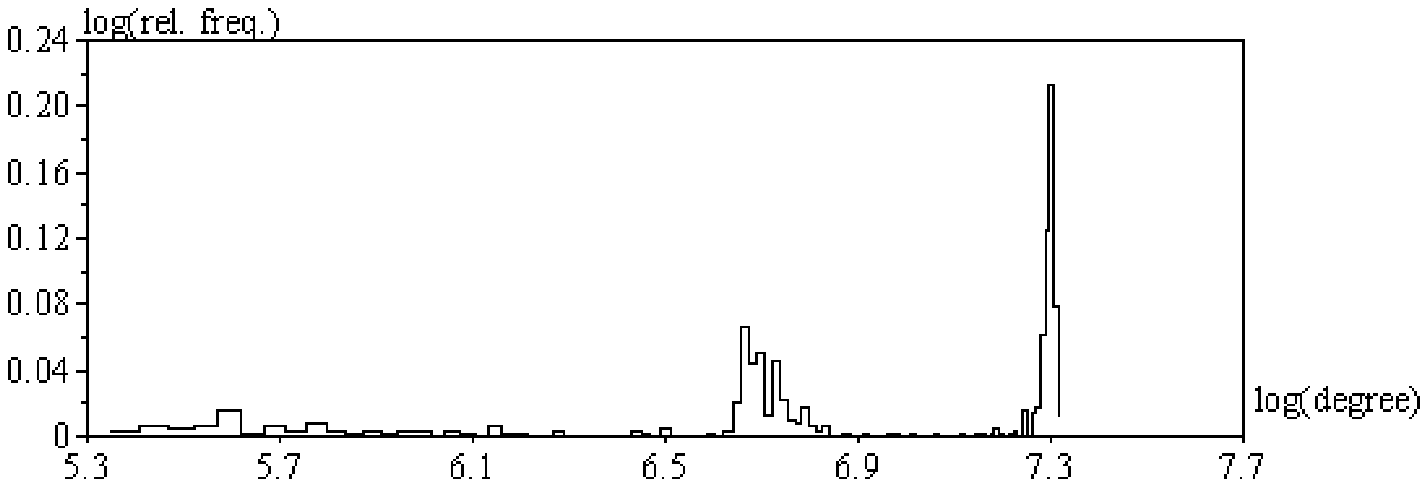} \\
   (e) \\
   \includegraphics[scale=.55,angle=-90]{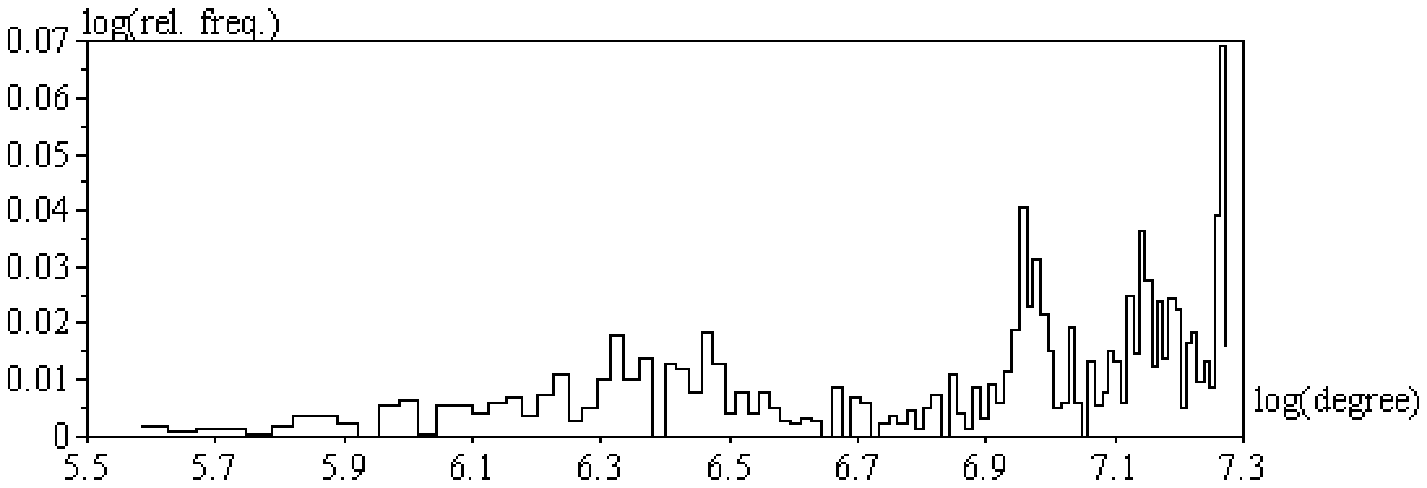} \\
   (f) \\   

   \caption{Dilog representation of the density of node degrees,
   considering the edge weights, for the networks derived from the
   images in Figures~\ref{fig:Lapis}(a) and
   Figures~\ref{fig:Lenna}(a).  The densities obtained from the
   respective adjacency matrix $A$, (c) and (d), and $A2$, (e) and
   (f).  The adopted threshold value was $T=0.1$.~\label{fig:dilog}}
\end{center}
\end{figure}

\section{Image Characterization in Terms of Topological Measurements} 
\label{sec:charact}

The possibility of using complex network concepts for image
characterization is illustrated in the following with respect to the
images in Figures~\ref{fig:Lapis}(a) and~\ref{fig:Lenna}(a).
Figures~\ref{fig:dilog}(a) and (b) presents the dilogarithm
representation of the node degrees, considering the respective
weights, for the network obtained for those two images considering
$T=0.1$.  The degree of a node $i$ was obtained by adding the weights
of all edges connected to that node.  Figures~\ref{fig:dilog}(c) and
(d) show the node degree densities obtained for the respective
adjacency matrices, while the node degrees considering
\emph{2-}expansions of the original networks are shown in
Figures~\ref{fig:dilog}(e) and (f).  Interestingly, the density
obtained for the image in Figure~\ref{fig:dilog}(a) presents two
peaks, suggesting the existence of two types of interactions (weak and
strong) between the image elements in the respective image.  The
densities obtained for the adjacency matrices $A$ indicate a
predominance of higher node degrees, indicating that the image
elements tend to be connected by many edges. The effect of the
\emph{2-}expansions in enhancing the connections between pixels can be
readily observed by comparing Figures~\ref{fig:dilog}(e)-(f) to
(c)-(d). It is clear from this example that rather different node
degree signatures are obtained for the two considered images,
suggesting that such features can be effective for image
characterization and classification.

\section{Hub-Based Image Segmentation}

\begin{figure}
 \begin{center} 
   \includegraphics[scale=.5,angle=-90]{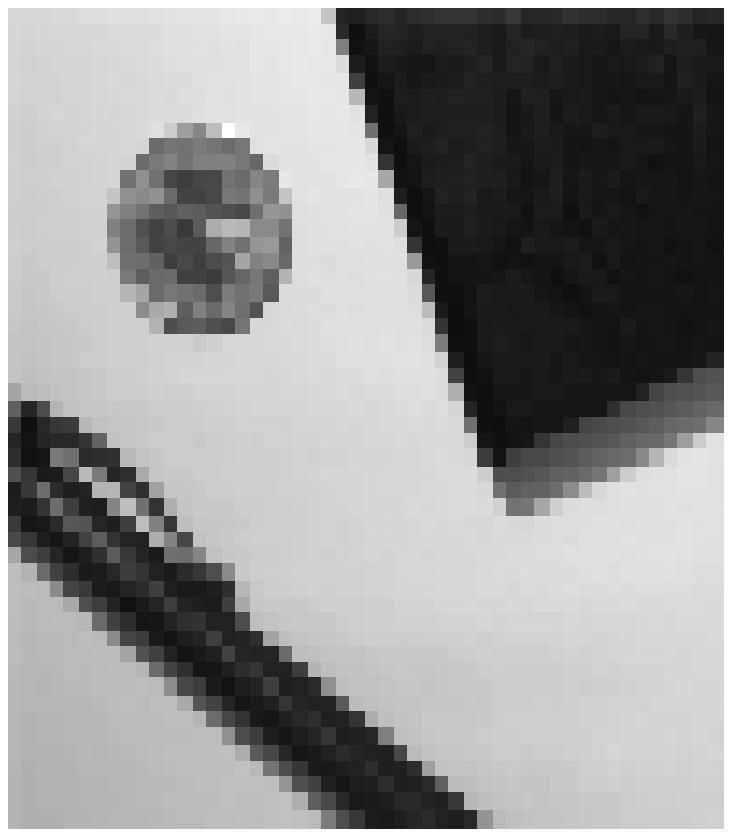} \hspace{0.5cm}
   \includegraphics[scale=.5,angle=-90]{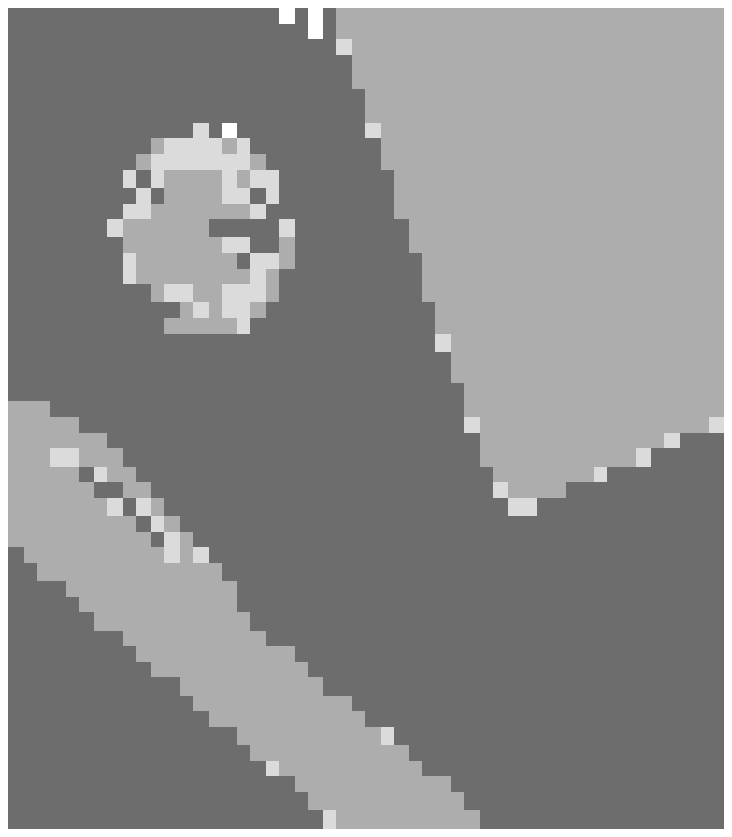} \\
   (a)  \hspace{4cm} (b) \\   
   \caption{Image containing a coin, a pen and part of a floppy disk
   (a) and respective segmentation (b) obtained for 
   $T=0.05$.~\label{fig:Lapis}} 
\end{center}
\end{figure}

Now we turn our attention to the important and challenging problem of
image segmentation.  One of the possible manners to approach such an
issue is by considering that each segmented region can be understood
as a connected component of the respective network whose nodes are
characterized by similar visual properties, such as the gray-level
relative uniformity inside the objects in Figure~\ref{fig:Lapis}(a).
Considering that the network \emph{hubs} are associated to a large
number of other nodes, it would be reasonable to expect that those
nodes share similar properties between themselves and also with the
respective hub.  This fact suggests the following algorithm for
partitioning the network into connected components:

\begin{tabbing}

Obtain the weight matrix $W$ from the image; \\
Threshold $W$ so as to obtain the adjacency matrix $A$; \\
For \=$k=1$ to $K$ do \\
  \> Iden\=tify the node $h$ with the highest node degree \\
  \>  \> in $A$ (the \emph{hub}); \\
  \> Join all nodes connected to $h$, and also this node, \\
  \>  \>into a detected cluster $C$; \\
  \> Remove from the adjacency matrix all edges \\
  \>  \>between the nodes in $C$;

\end{tabbing}

\begin{figure}
 \begin{center} 
   \includegraphics[scale=.5,angle=-90]{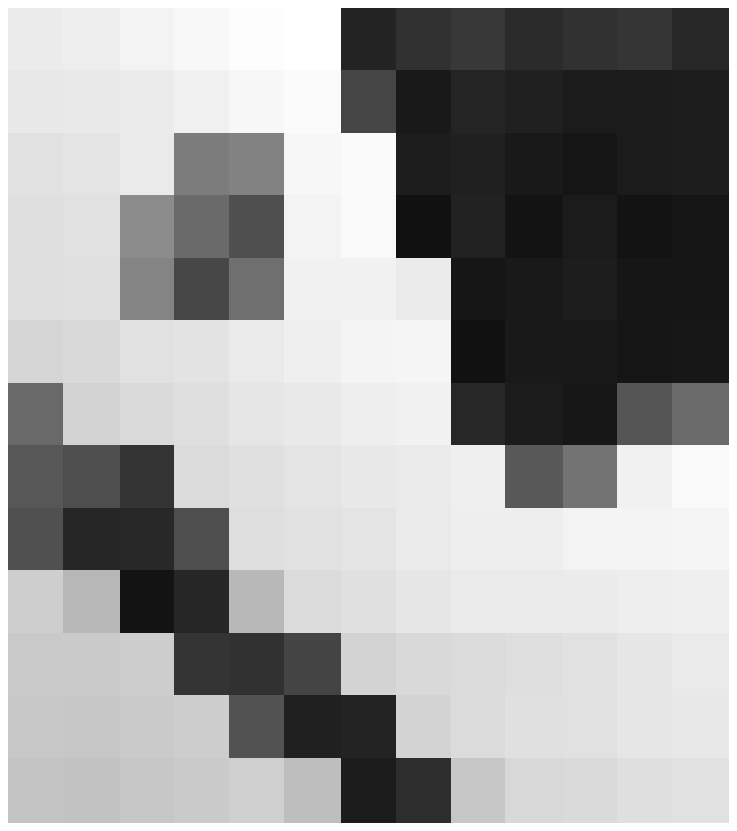} \hspace{0.5cm} 
   \includegraphics[scale=.5,angle=-90]{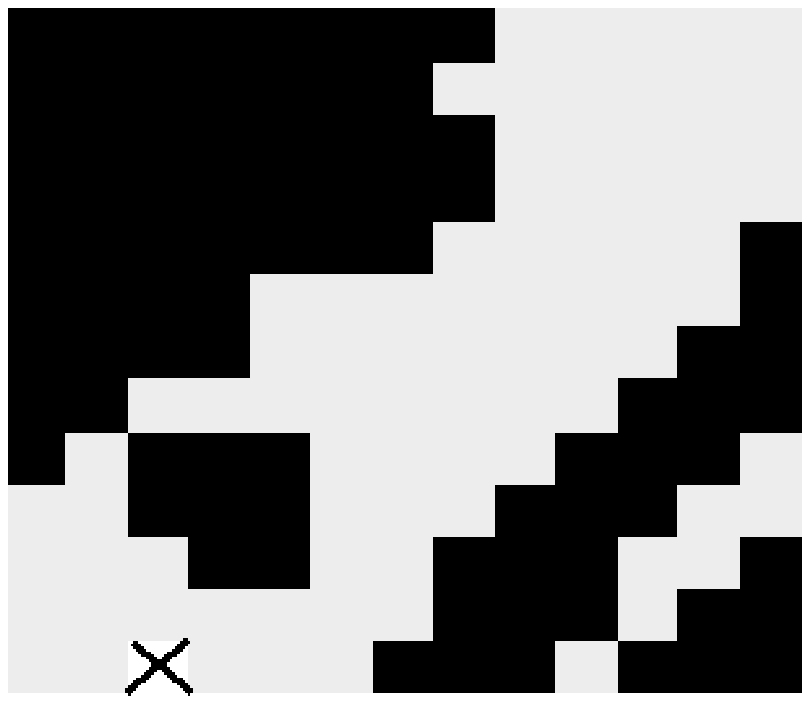} \\
   (a) \hspace{4cm} (b) \\
   \includegraphics[scale=.5,angle=-90]{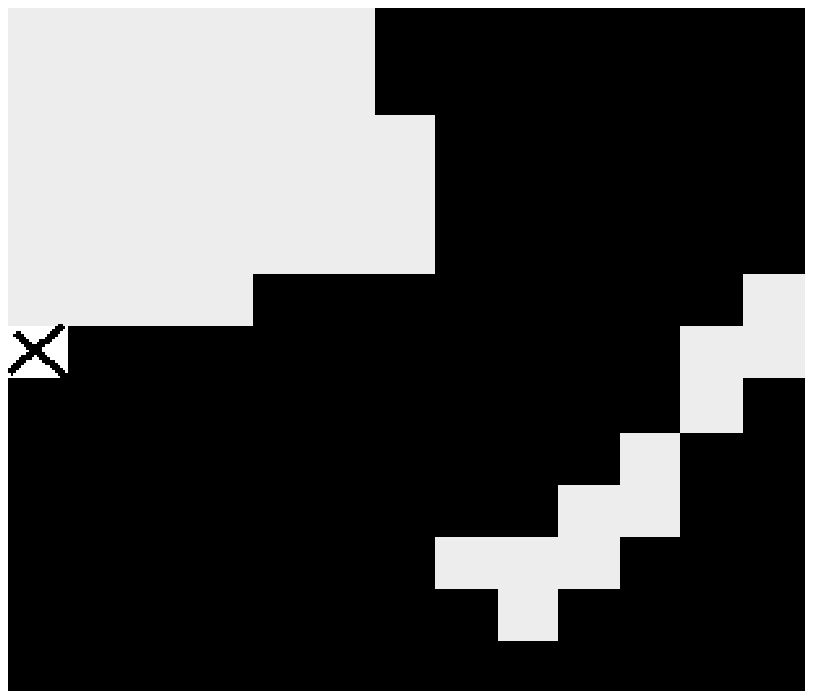} \hspace{0.5cm}
   \includegraphics[scale=.5,angle=-90]{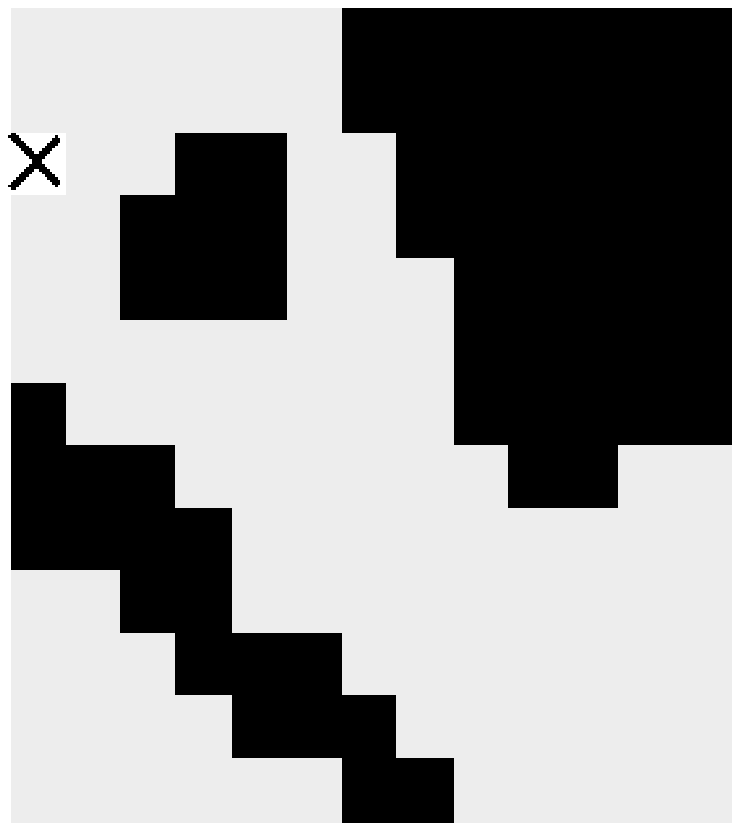} \\ 
   (c) \hspace{4cm} (d) \\ 
   \includegraphics[scale=.5,angle=-90]{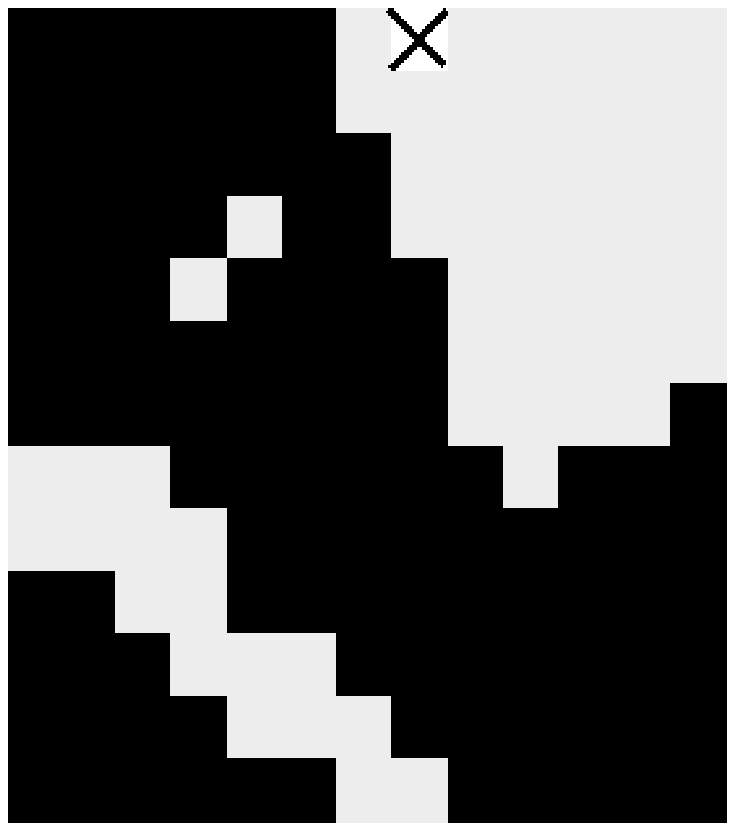} \hspace{0.5cm} 
   \includegraphics[scale=.5,angle=-90]{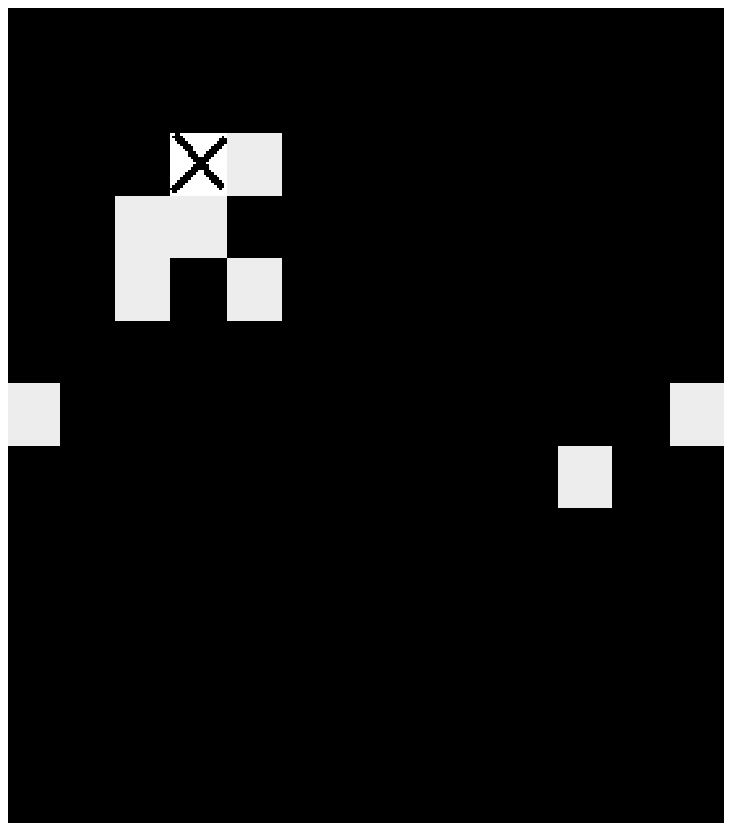} \\
   (e) \hspace{4cm} (f) \\ 

   \caption{A coarse gray-level image (a) and the first two hubs,
   marked by crosses in (b)-(c), and respective clusters (light
   gray-level) obtained by considering traditional edges.  The three
   identified hubs (d)-(f) and respective clusters obtained from the
   hyperedges defined by the \emph{2-}expansions.~\label{fig:ex_hub}}
\end{center}
\end{figure}

The number of repetitions $K$ corresponds to the number of expected
clusters.  This stop condition can be modified in order to consider
the number of nodes obtained in each cluster, terminating the
algorithm when all pixels are labeled or when the remaining number of
unlabelled pixels is smaller than a threshold value.  While the
weighted edges in the networks representing images provide an
indication about the dissimilarity degree between pairs of pixels,
they have been experimentally verified to lead to unsuitable clusters.
This is illustrated in Figure~\ref{fig:ex_hub}(b)-(c).  The cluster
obtained for the first identified hub, marked by the cross in
Figure~\ref{fig:ex_hub}(b), was unsuitable in the sense that it missed
part of the background (the black region at the bottom right corner of
the image).  The cluster obtained for the second dominant hub, shown
in Figure ~\ref{fig:ex_hub}(c) also missed part of the pen (the
diagonal black stripe).  This tendency to produce incomplete clusters
is related to the fact that there are several pixels indirectly
connected to the respective hub which, though sharing visual
properties, are not included into the cluster.  This problem can be
addressed by considering the \emph{2}-expansions of the network, which
will connect all pixels united to the hub by a \emph{2-}path.  At the
same time as the above problem is reduced, pixels directly connected
to the hub which are not also connected through at leas one
\emph{2-}path are ignored by the expansion.  This enhanced complex
network based segmentation algorithm therefore is given as follows

\begin{tabbing}

Obtain the weight matrix $W$ from the image; \\
Threshold $W$ so as to obtain the adjacency matrix $A$; \\
Obta\=in the adjacency matrix $A2$ for the \emph{2-}expanded \\
   \> network \\
For $k=1$ to $K$ do \\
  \> Iden\=tify the node $h$ with the highest node degree \\
  \>  \> in $A2$ (the \emph{hub}); \\
  \> Join all nodes connected to $h$, and also this node, \\
  \>  \>into a detected cluster $C$; \\
  \> Remove from the adjacency matrix all edges \\
  \>  \>between the nodes in $C$;

\end{tabbing}

\begin{figure}
 \begin{center} 
   \includegraphics[scale=.5,angle=-90]{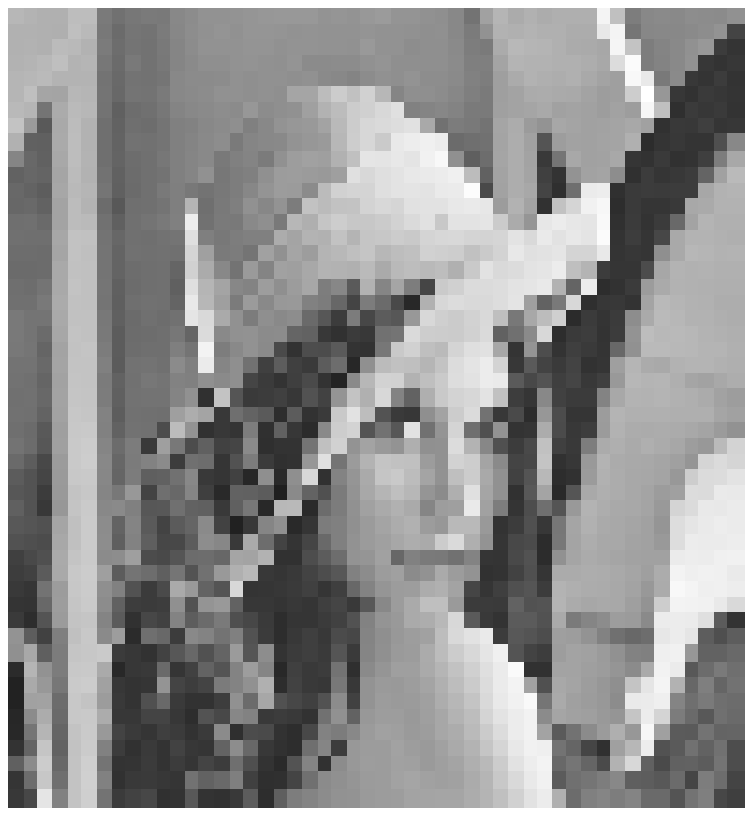} \hspace{0.5cm}
   \includegraphics[scale=.5,angle=-90]{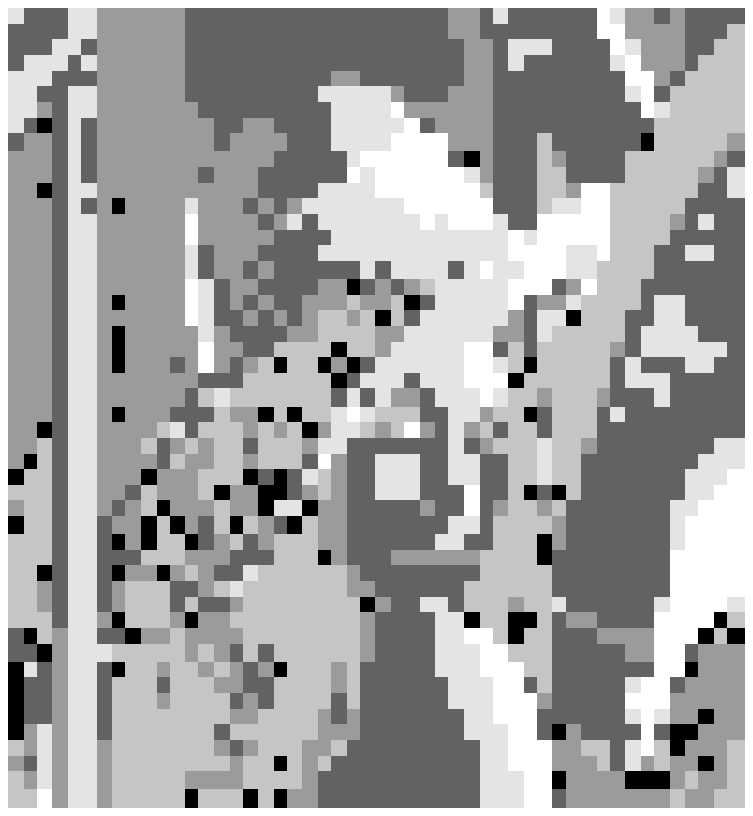} \\
   (a)  \hspace{4cm} (b) \\   
   \caption{Lenna's image (a) and respective segmentation 
   (b) obtained for $T=0.05$.~\label{fig:Lenna}} 
\end{center}
\end{figure}

Figure ~\ref{fig:ex_hub} illustrates the potential of the enhanced
segmentation algorithm with respect to the initial image (a), which
corresponds to a coarse version of the image in
Figure~\ref{fig:Lapis}(a), which is considered here for the sake of
illustration.  The first three identified hubs (crossed pixel) and
respective clusters (light gray-level values) are shown in (d)-(f).
The clusters obtained for the two first hubs, shown in
Figures~\ref{fig:ex_hub}(d) and (e) resulted considerably superior to
those obtained by using the traditional edges (i.e. the clusters in
Figures~\ref{fig:ex_hub}(b) and (c)).  Interestingly, the first
identified hub, which correspond to the image background, implied in a
respective segmented region (light gray-level) merging the two
portions of the background which were previously separated. Further
examples of the potential of the proposed segmentation algorithm are
given in Figures~\ref{fig:Lapis}(b) and~\ref{fig:Lenna}.

\begin{figure}
 \begin{center} 
   \includegraphics[scale=.55,angle=-90]{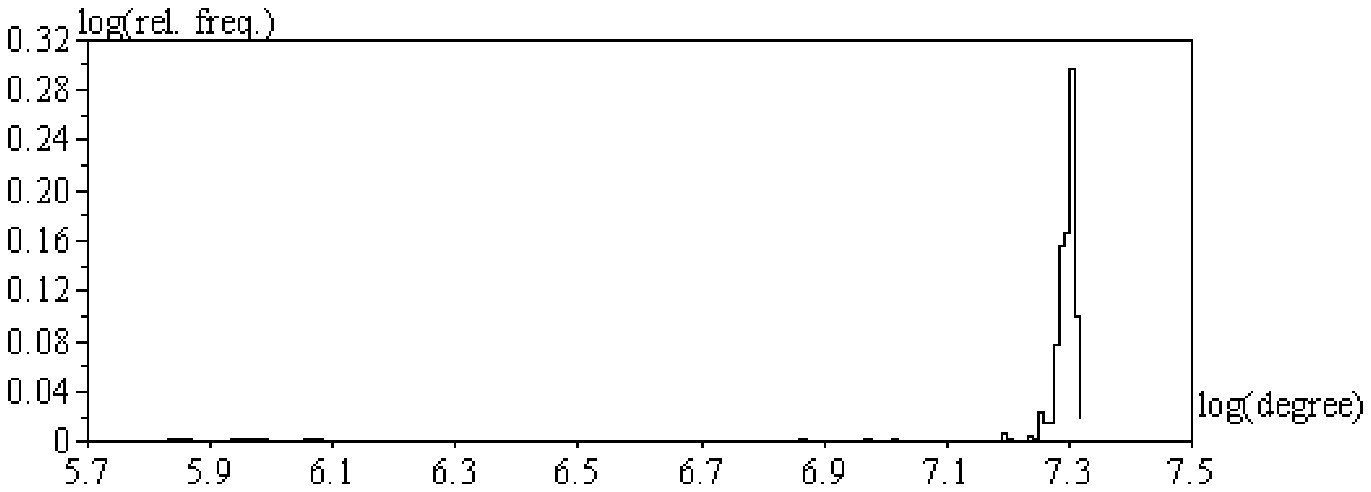} \\
   (a) \\
   \includegraphics[scale=.55,angle=-90]{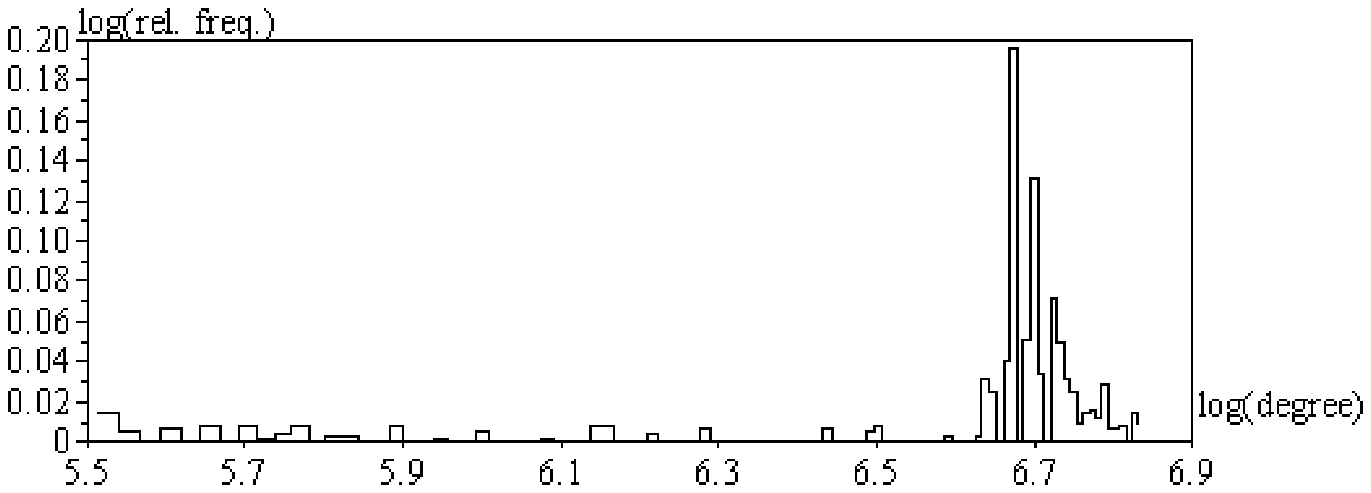} \\
   (b) \\   

   \caption{The node degree densities obtained from the adjacency
   matrix (after \emph{2-}expansion) describing the pixel connections
   inside the first (a) and second (b) obtained segmented regions
   considering the image in
   Figure~\ref{fig:Lapis}(a).~\label{fig:partial}}
\end{center}
\end{figure}

In similar fashion to the characterization of whole images in terms of
complex network measurements proposed in Section~\ref{sec:charact}, it
is possible to apply such an approach to characterize the regions
produced by segmentation algorithms such as that described above.
This can be immediately accomplished from adjacency matrices defined
for each segmented region.  Figure~\ref{fig:partial} illustrates the
node degree densities (after \emph{2-}expansion) obtained for the
segmented regions (light gray-level) in Figures~\ref{fig:ex_hub}(d)
and (e).  It is clear from this example that rather distinct
connecting patterns can be obtained for two different regions of the
same image.  Such region-oriented topological measurements obtained
for the representing network provide valuable subsidies for object
recognition and texture characterization.

\section{Concluding Remarks}

This paper has illustrated how digital images can be effectively
represented, characterized and analyzed in terms of graphs taking into
account several ranges of spatial interactions.  The potential of such
a framework has been illustrated for image characterization in terms
of node degree densities and for image segmentation by using a
combination of the concepts of hub and \emph{2-}expansions of a
network.  Several future developments have been motivated by such
results.  First, it would be interesting to investigate how complex
network measurements such as the node degree densities, clustering
coefficient and minimal length can be used for the characterization of
broad classes of images (e.g. landscale scenes, textures, medical
images, etc.).  Another promising possibility is to use higher
\emph{L-}expansions in order to further enhance the proposed
segmentation approach.  At the same time, the concept of hub-oriented
clustering considered in this work can be immediately extended for
general pattern recognition, allowing the identification of prototypes
(i.e. the respective hubs) of each identified class.  Along similar
lines, it would be interesting to investigate how alternative
partitioning schemes such as that reported in \cite{Holme:2003} behave
for image segmentation, with prospects for integrating approaches.
Other further developments include the consideration of the segmented
regions as spatial networks \cite{Costa_Vor:2003} and the
topographical integration of networks obtained for distinct image
properties \cite{Costa_Topo:2003}.  Finally, it would be particularly
interesting to check how the topological connections of the networks
representing digital images change with the progressive increase of
the threshold $T$, with the possible occurrence of critical phenomena
(e.g. percolation).

\begin{acknowledgments}

The author is grateful to FAPESP (proc. 99/12765-2), CNPq
(proc. 301422/92-3) and Human Frontier Science Program for financial
support.

\end{acknowledgments}

%\clearpage
 
%\bibliographystyle{plain}
\bibliography{complima}

\end{document}